\documentclass[a4paper]{jpconf}
\usepackage{amsmath} 
\usepackage{amsfonts}
\usepackage{amssymb}
\usepackage{bm} 
\usepackage[colorlinks]{hyperref}

\usepackage{graphicx}
\begin{document}
\title{Two-parameter boundary condition applied to transverse acoustic impedance of a Fermi liquid}

\author{J. A. Kuorelahti and E. V. Thuneberg}

\address{Nano and molecular systems research unit, University of Oulu, Finland}


\begin{abstract}
We discuss surface scattering in terms of bidirectional reflectance distribution function. We formulate a boundary condition that in addition to diffuse and specular reflection has tilted diffuse reflection. We apply the boundary condition to calculate the transverse acoustic impedance of a normal-state Fermi liquid. Some additional results on the impedance in a slab geometry are presented.
\end{abstract}

\section{Introduction}

In this article we are interested in the reflection of quasiparticles of a Fermi liquid from surfaces. Very often, only diffuse or specular reflection, or their combination, has been considered \cite{Buchholtz81,Zhang87,Vorontsov07,Nagai08,Thuneberg11,Mizushima16}.
The purpose is to test a boundary condition that goes beyond these. The reflectance of electromagnetic waves from a surface is a topic of long history \cite{Beckmann63}. The reflectance is commonly specified by {\em bidirectional reflectance distribution function} (BRDF) \cite{Nicodemus77}. Especially in the connection of computer graphics, there is a need for a simple but physically realistic rendering, and several different reflectance models have been proposed, see e.g.\ Refs.\ \cite{Ward92,Neumann99,Geisler-Moroder10}.
We formulate a general boundary using BRDF and propose a simple generalization of the diffuse-specular boundary condition that has a tilt in the diffusive part. We apply the boundary condition to a normal-state Fermi liquid film on a transversely oscillation plane \cite{Kuorelahti16}. Besides presenting results of the new boundary condition, some additional results are shown.

\section{Bidirectional reflectance distribution function}

Let us consider elastic scattering from a surface. We study an approximately planar piece of the surface, which has area $A$ and surface normal $\hat{\bm n}$. The amount of both incoming and reflected monochromatic radiation in a given direction can be described by radiance. Radiance $L$ is the flux $\Phi$ of the radiation arriving to a surface area $A$ from a small solid angle $\Omega$ around the incoming direction $\hat{\bm p}$ per solid angle and per area projected in the direction of incidence. As a formula,
$L=\Phi/\Omega A \left|\cos\theta\right|$, where $\cos\theta=\hat{\bm n}\cdot\hat{\bm p}$.  The same formula describes also the radiance  reflected from the surface element to a solid angle around the outgoing direction.
The reflected radiance $L_{r}(\hat{\bm p})$ is given in terms of the incoming radiance $L_{i}(\hat{\bm p})$ as
\begin{eqnarray}
L_{r}(\hat{\bm p})
=\int_{\rm in}d\Omega' \left|\cos\theta'\right| f(\hat{\bm p},\hat{\bm p}')L_{i}(\hat{\bm p}'),
\label{e.bdrdf}\end{eqnarray}
where $f(\hat{\bm p},\hat{\bm p}')$ is the {\em bidirectional reflectance distribution function} (BRDF). Here $\int_{\rm in}d\Omega$ is the integral over the solid angle of incoming momenta, $\hat{\bm n}\cdot\hat{\bm p}=\cos\theta<0$. The factor $\left|\cos\theta\right|$ is included in (\ref{e.bdrdf}) to compensate the same factor appearing in the definition of $L_{i}$. For a perfectly reflecting surface the flux conservation then  sets 
\begin{eqnarray}
\int_{\rm in}d\Omega' \left|\cos\theta'\right|f(\hat{\bm p},\hat{\bm p}')=1.
\label{e.bdrdfc}\end{eqnarray}
Apparently $f$ has to be non-negative and time reversal invariance requires
\begin{eqnarray}
f(\hat{\bm p},\hat{\bm p}')=f(-\hat{\bm p}',-\hat{\bm p}).
\label{e.bdrdtr}\end{eqnarray}

It is useful to notice that $\int_{\rm in}d\Omega \left|\cos\theta\right|$ can be interpreted as an integral over {\em projected solid angle}. Thus it is equivalent to integration of over a unit disk. Writing in spherical coordinates
$\hat{\bm p}=(\hat{\bm x}\cos\phi+\hat{\bm y}\sin\phi)\sin\theta+\hat{\bm z}\cos\theta$, where $\hat{\bm z}=\hat{\bm n}$, and defining $\rho=\sin\theta$, the integral is 
\begin{eqnarray}
\int_{\rm in}d\Omega \left|\cos\theta\right|\ldots=\int_0^{2\pi}d\phi\int_0^1d\rho\,\rho\ldots,
\label{e.unidisc}\end{eqnarray}
i.e.\ integration over the unit disk.
It is also equivalent to an integral (or sum) over ``transverse modes''. This language is often used to describe electrical conductance in a ballistic channel, where each occupied mode contributes equally to the flux \cite{Datta}. In a quantum treatment, BRDF can be presented as a square of a unitary scattering matrix.

The most common model forms of BRDF are {\em diffuse reflection}
\begin{eqnarray}
f(\rho,\phi,\rho',\phi')=\frac1\pi
\label{e.bcdiff}\end{eqnarray}
and {\em specular reflection}
\begin{eqnarray}
f(\rho,\phi,\rho',\phi')=\frac1{\rho'}\delta(\rho-\rho')\delta(\phi-\phi').
\label{e.bcspec}\end{eqnarray}
Diffuse reflection means that the scattered distribution is independent of the incoming direction and is equally probable in any direction with the same projected solid angle. 
Specular reflection means that the incoming flux in direction $\hat{\bm p}'$ is all scattered in the direction $\hat{\bm p}=\hat{\bm p}'-2\hat{\bm n}(\hat{\bm n}\cdot\hat{\bm p}')$. A commonly used boundary condition is a combination of the diffuse and specular scattering,  
\begin{eqnarray}
f(\rho,\phi,\rho',\phi')=\frac{1-s}{\pi}+\frac{s}{\rho'}\delta(\rho-\rho')\delta(\phi-\phi')
\label{e.spbc}\end{eqnarray}
with parameter $s$ describing specularity, $0\le s\le 1$. 

Our plan is to look for a more general boundary condition than the combination of diffuse  and specular  scattering (\ref{e.spbc}). One possibility is to spread the specular scattering $\delta$ functions to have a finite width. This is especially used in computer graphics \cite{Ward92,Neumann99,Geisler-Moroder10}. However, we did not find a simple model that would satisfy the conditions (\ref{e.bdrdfc}) and (\ref{e.bdrdtr}) precisely. Therefore we turned our attention to the generalization of the diffuse limit. A general $f$ could be presented in terms of  Zernike polynomials. They are an orthogonal set of functions defined on the unit disc (\ref{e.unidisc}).
Instead of formulating such expansion in general, we limit here to the leading order generalization of (\ref{e.bcdiff}), and we write our two-parameter boundary condition
\begin{eqnarray}
f(\rho,\phi,\rho',\phi')=\frac{1-s}{\pi}+a\rho\rho'\cos(\phi-\phi')
+\frac{s}{\rho'}\delta(\rho-\rho')\delta(\phi-\phi').
\label{e.shf1}\end{eqnarray}
In order to ensure non-negativity, we require $|a|\le(1-s)/\pi$. The new middle term describes tilt of the reflected distribution in the direction of in-plane component of the incoming momentum.

\section{Transverse acoustic impedance of a Fermi liquid}

We consider a degenerate Fermi system as described by Landau's theory \cite{Landau57}.  As all quasiparticles effectively propagate at the Fermi velocity $v_F$, we can apply the general boundary condition (\ref{e.bdrdf}) directly to the quasiparticle distribution function.  We use the energy-integrated distribution $\psi_{\hat{\bm p}}$, which gives the local energy shift of the Fermi surface relative to the equilibrium Fermi level \cite{Thuneberg11,Kuorelahti16}. Taking into account that the surface is moving with velocity $\bm u$, we have to replace $\psi_{\hat{\bm p}}\rightarrow \psi_{\hat{\bm p}}-p_F\hat{\bm p}\cdot\bm u$ \cite{Thuneberg11}. The  boundary condition (\ref{e.bdrdf}) takes the form
\begin{eqnarray}
\psi_{\hat{\bm p}}=\int_{\rm in} d\Omega'\left|\cos\theta'\right| f(\hat{\bm p},\hat{\bm p}')\psi_{\hat{\bm p}'}+p_F\bm u\cdot\left(\hat{\bm p}-\int_{\rm in} d\Omega'\left|\cos\theta'\right|f(\hat{\bm p},\hat{\bm p}')\hat{\bm p}'\right)
\end{eqnarray}
for $\hat{\bm p}\cdot\hat{\bm n}>0$.
For transverse impedance with $\bm u=u\hat{\bm x}$, the distribution takes the form $\psi_{\hat{\bm p}}=\cos\phi\sin\theta\,\psi(\cos\theta,\zeta)$, where $\zeta$ is the dimensionless vertical coordinate in the liquid \cite{Kuorelahti16}. The BRDF (\ref{e.shf1}) then leads to the boundary condition
\begin{eqnarray}
\psi(\mu>0,0)=\pi a\int_{-1}^0d\mu'\, |\mu'|(1-\mu'^2)\psi(\mu',0)+s\psi(-\mu,0)+\left(1-\textstyle{\frac{\pi}{4}}a-s\right)p_Fu.
\end{eqnarray}
Here the last term appears as the driving term in the calculation of the transverse acoustic impedance. We can define {\em  momentum transfer efficiency} $q=1-\frac{\pi}{4}a-s$. We can expect that the main effect of changing $a$ and $s$ is through the change of $q$. Then we can study the remaining effect of $s$ at a constant $q$.

The calculation of the transverse acoustic impedance is described in Ref.\ \cite{Kuorelahti16}. The reflection on the lower, oscillating plane is described by parameters $q$ and $s$. The upper surface can be free ($q_2=0$, $s_2=1$) or a diffusely reflecting, stationary surface ($q_2=1$, $s_2=0$). In addition, the calculation depends on the following dimensionless parameters: the ratio of the mean free path $l$ to the thickness of the film $d$,  $\Omega=\omega d/v_F(1+\frac13F_1)$ containing the angular frequency $\omega$ of the oscillation, the Fermi liquid parameters $F_1$ and $F_2$, and parameter $\xi_2$ describing the anisotropy of the quasiparticle relaxation rate. In the figures we study the effect of $q$ and $s$ as well as other parameters on the acoustic impedance $Z=Z'+iZ''$. $Z$ is expressed in units of $p_F n$ or  $p_F n\Omega$, where $p_F$ is the Fermi momentum and $n$ the fermion number density.

\begin{figure}[tb] 
    \centering
    \includegraphics[width=0.9\linewidth]{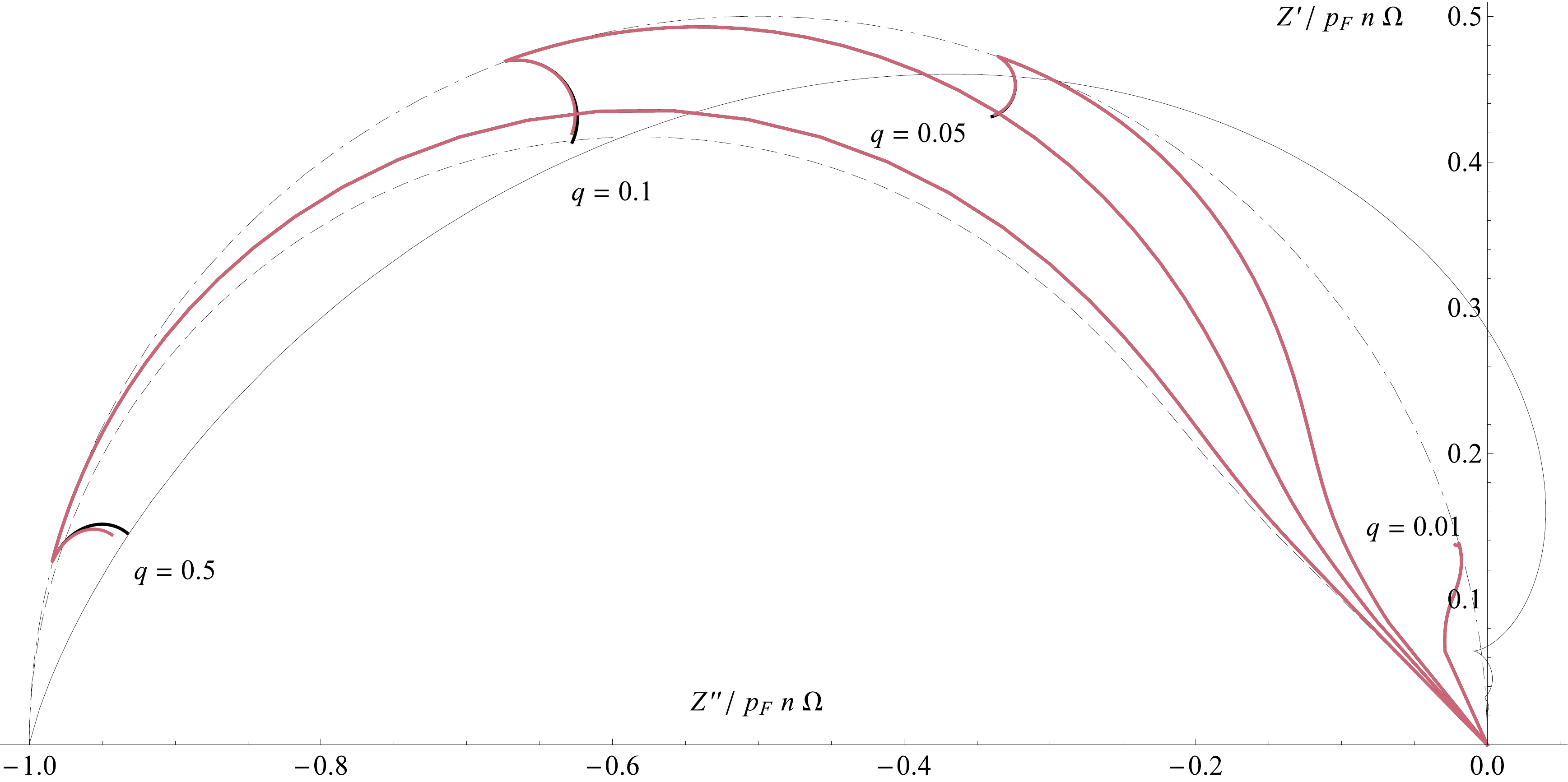} 
    \caption{The acoustic impedance $Z$ of a Fermi liquid film with a free upper surface: the real and imaginary parts as a parametric plot with parameter $l/d$ \cite{Kuorelahti16}. The red/black curves are labelled by the momentum transfer efficiency $q=0.5$, 0.1, 0.05 and 0.01. There is very little difference between curves with $s=0$ (red) and $s=1-q$ (black, visible only if different from $s=0$ case). Other parameters are $\Omega=0.01$, $F_1=6$, $F_2=0$, and $\xi_2=0.35$. For comparison, the figure shows the single-body limit (dash-dotted), the hydrodynamic limit (dashed), and the ballistic gas limit (thin solid line). }
    \label{f.smallOmega_s_NB}
 \end{figure}

\begin{figure}[tb] 
    \centering
    \includegraphics[width=0.8\linewidth]{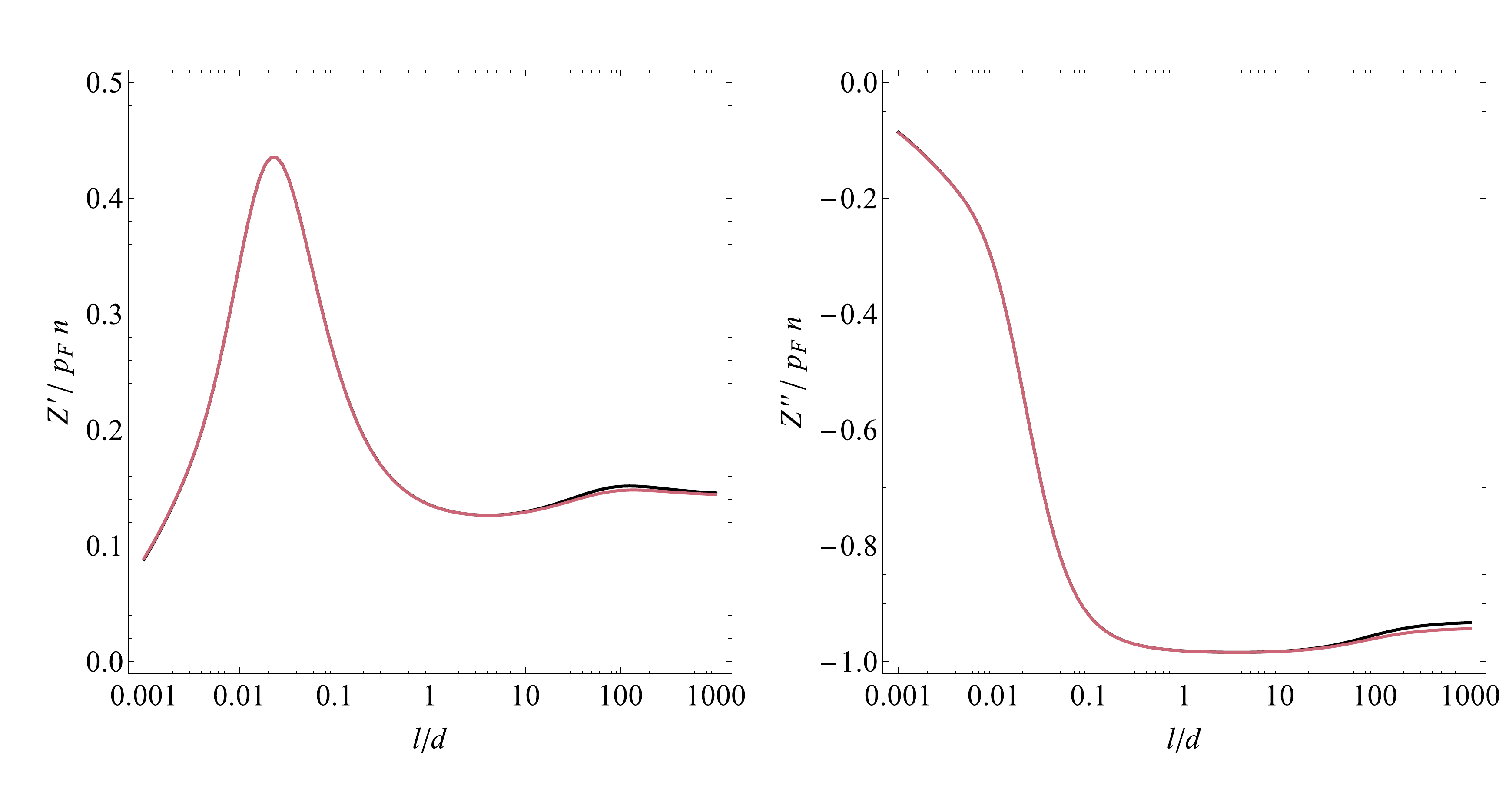} 
    \caption{The real and imaginary parts of $Z$ of a Fermi liquid film with a free upper surface as a function of $l/d$. The curves and parameters are the same as for $q=0.5$ in Fig.\ \ref{f.smallOmega_s_NB}.}
    \label{f.smallOmega_NB_ld}
 \end{figure}

Figures \ref{f.smallOmega_s_NB} and \ref{f.smallOmega_NB_ld} display the case of small $\Omega$, meaning a low frequency or a thin film. We see that the new boundary condition has an effect only in the ballistic region $l/d>1$, and even there the difference is small. Besides the hydrodynamic limit curve and ballistic limit points,  Fig.\ \ref{f.smallOmega_s_NB}  displays  a half circle  of radius 0.5 centered at $-0.5i$ (dash-dotted line). This curve is obtained if the liquid would  be considered as a single rigid body that is dissipatively coupled to the substrate. We see that parts of the curves closely follow this ``single body model'' in the regime $l/d\lesssim1$. We have checked that the flow profile in these regions indeed is like motion as a rigid body that slips from the motion of the substrate. 

A model similar to the single-body model was used in Ref.\ \cite{Casey04} to fit experimental data on the decoupling of liquid $^3$He from a torsional oscillator. Both models give the same parametric curve, but the temperature dependences are different. The experimental finding in Ref.\ \cite{Casey04} was an increasing  $Z''$ with decreasing temperature. In our model the temperature dependence is visible in Fig.\ \ref{f.smallOmega_NB_ld} as $l/d$ increases with decreasing temperature. We see mainly decreasing $Z''$ and only a slight increase for  $l/d\gg1$. We see that the new boundary condition (\ref{e.shf1}) gives similar result as the earlier one (\ref{e.spbc}).  The reasons for the difference between the experiment and our Fermi-liquid model remain open.

\begin{figure}[tb] 
    \centering
    \includegraphics[width=0.95\linewidth]{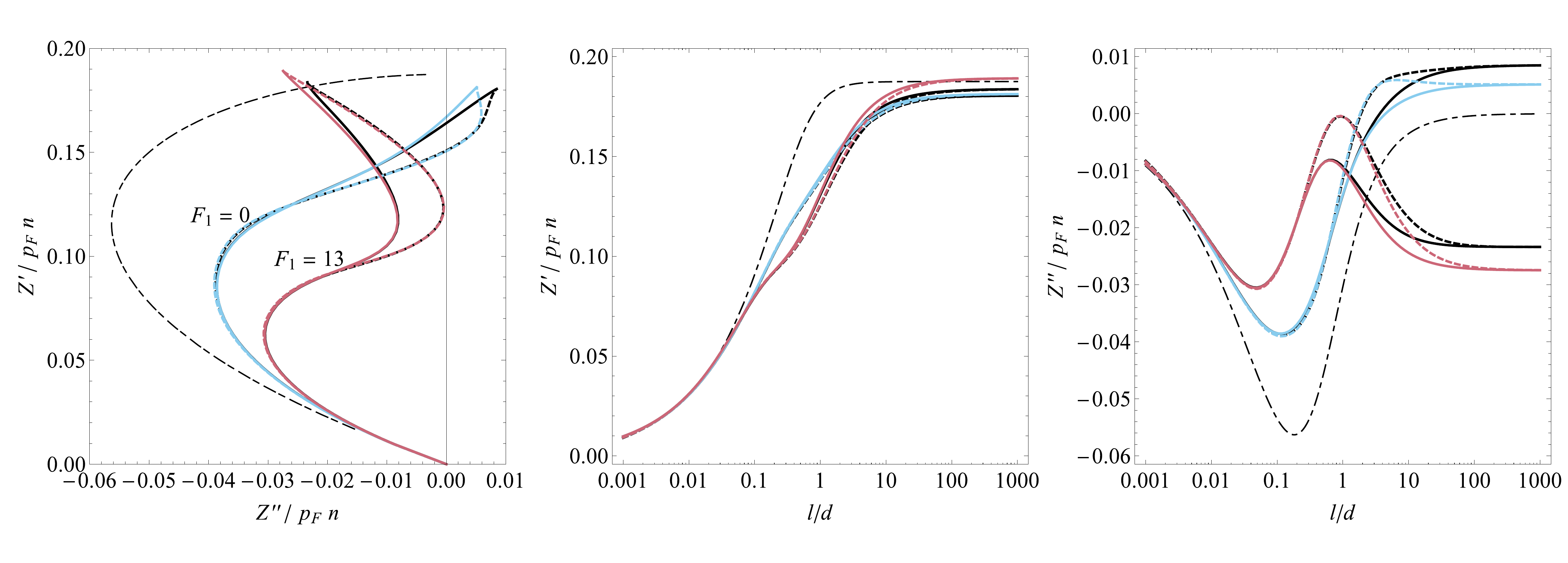} 
    \caption{The acoustic impedance $Z$ of a Fermi liquid film with a free upper surface. Displayed are, from left to right, the real and imaginary parts as a parametric plot, the real part and the imaginary part as functions of $l/d$. The curves are for $s=0$ (red and blue) and $s=0.25$ (black), $\xi_2=0.35$ (dashed) and $\xi_2=1$ (solid), $F_1=0$ (blue and black) and $F_1=13$ (red and black). Other parameters are $q=0.75$, $\Omega=1$, and $F_2=0$. For comparison, the thick-film limit at $q=1$, $F_1=0$, $\xi_2=1$ is shown by dash-dotted line.}
    \label{f.Xi2_F1_and_NB}
 \end{figure}

Figure \ref{f.Xi2_F1_and_NB} gives results for $\Omega=1$, which could be relevant for experiments at higher frequencies. We see that there is a small effect of $s$ at a given $q$ in the ballistic region. The same figure also compares the effect of the relaxation anisotropy parameter $\xi_2$. It has effect particularly on $Z''$ but the effect vanishes in the ballistic limit, as it should be.

\begin{figure}[tb] 
    \centering
    \includegraphics[width=0.95\linewidth]{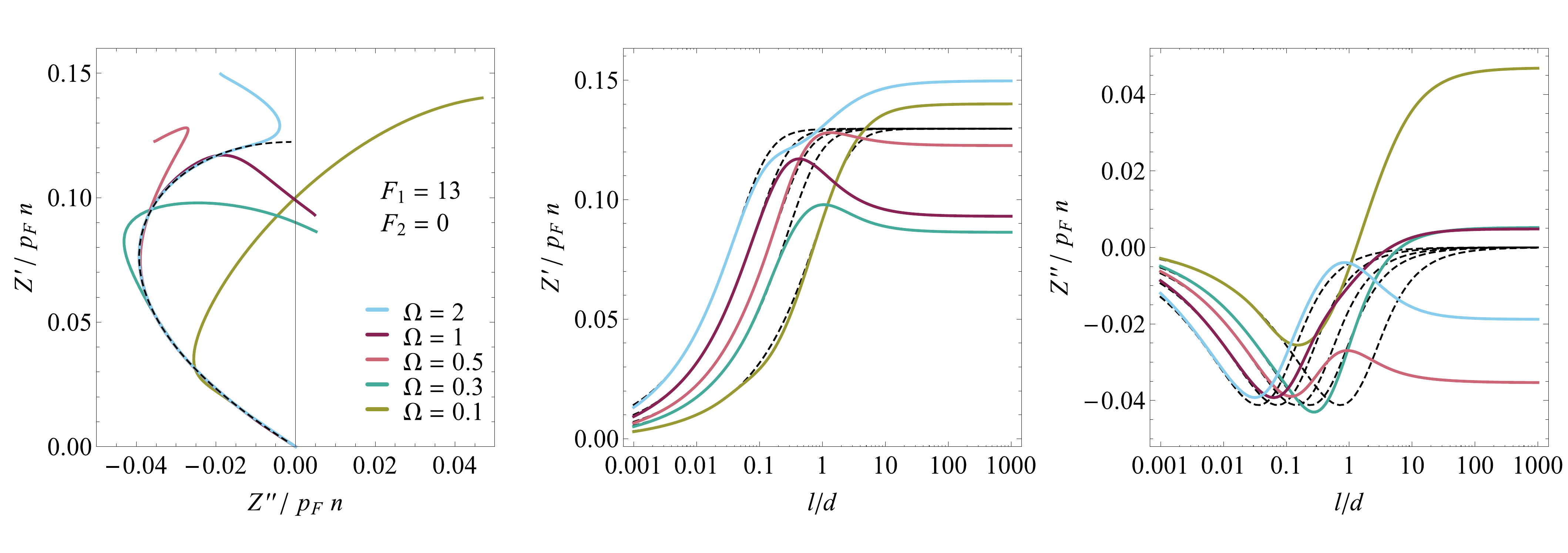} 
     \includegraphics[width=0.95\linewidth]{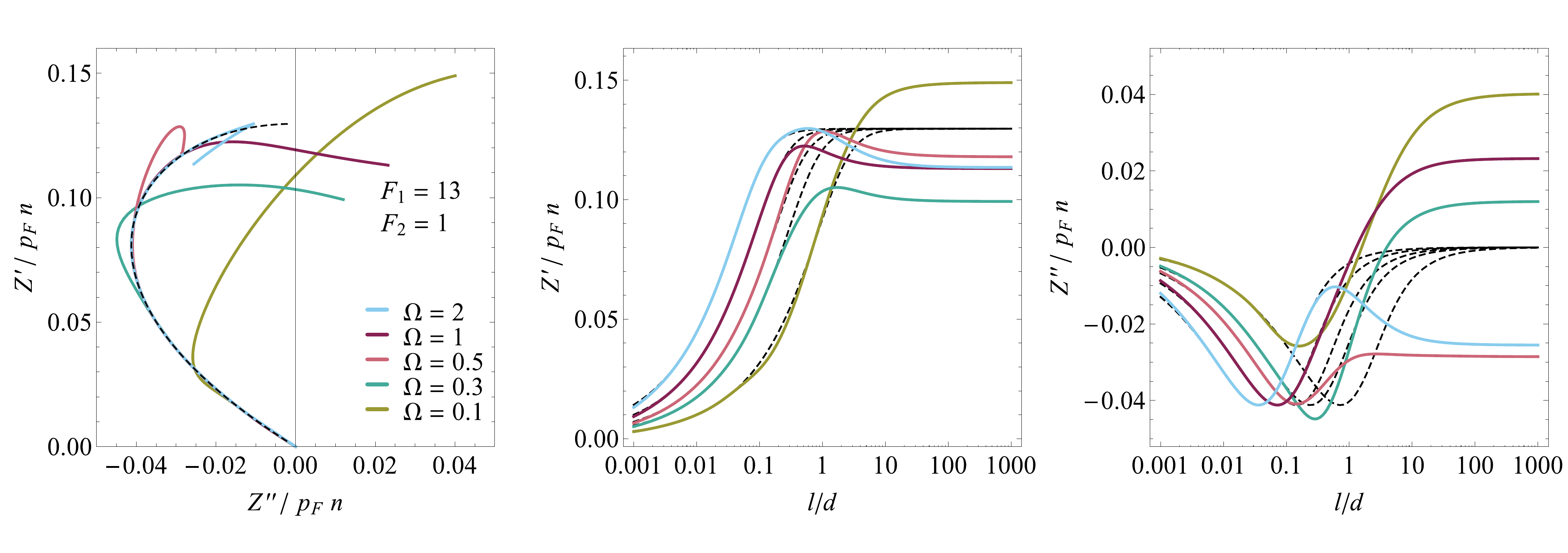} 
    \caption{The acoustic impedance $Z$ of a Fermi liquid film with a stationary, diffuse upper surface. The thick-film limit is shown by a dashed line. The parameters are $q=1$, $s=0$, $\xi_2=0.35$.}
    \label{f.nonscaledF20}
 \end{figure}

 Figure \ref{f.nonscaledF20} studies the case of a Fermi liquid in the space between a stationary and an oscillating surface, which both are diffusely reflecting. The shown results complement the special cases studied in Ref.\ \cite{Kuorelahti16}.
 
 \section{Conclusions}

 We have suggested a boundary condition that goes beyond the often used combination of specular and diffuse reflection. In application to transverse acoustic impedance of a Fermi liquid, we see that at least in the cases studied, the boundary condition has only minor effect at constant momentum transfer efficiency of the reflection.

\ack
We thank Jenny and Antti Wihuri foundation, Oskar \"Oflunds Stiftelse sr, the Academy of Finland and Tauno T\"onning foundation for financial support.

\end{document}